\documentclass[12pt,a4paper]{article}
\usepackage{cite}
\usepackage{graphics}
\usepackage{amssymb}

\begin{document}

\title{An improved Rosenbluth Monte Carlo scheme for cluster counting and lattice
animal enumeration  } \author{ C M  Care \thanks{Materials Research Institute,
Sheffield Hallam University, Pond Street, Sheffield, S1 1WB, UK}  \and R Ettelaie
\thanks{Colloids and Rheology Unit, ICI Wilton, PO Box 90, Wilton, Middlesbrough,
Cleveland, TS90 8JE, UK}  } \maketitle

\begin{abstract}

We describe an algorithm for the Rosenbluth Monte Carlo enumeration of clusters and
lattice animals. The method may also be used to calculate associated properties such
as moments or perimeter multiplicities of the clusters.  The new scheme is an
extension of the Rosenbluth method for growing polymer chains and is a simplification
of a scheme reported earlier by one of the authors.  The algorithm may be used to
estimate the number of  distinct lattice animals on any  lattice topology. The method
is validated against exact and Monte Carlo enumerations for clusters up to size 50, on
a two dimensional square lattice and three dimensional simple cubic lattice.  The
method may be readily adapted to yield Boltzmann weighted averages over clusters.

\label{abstract}
\end{abstract}

\section{Introduction}

The enumeration of lattice animals is an important problem in a variety of physical
problems including nucleation \cite{jacucci:83a}, percolation \cite{edwards:92a} and
branched polymers \cite{ball:96a}. A lattice animal is a cluster of  $N$ connected
sites on a lattice with given symmetry and dimensionality and we seek to enumerate all
distinct animals with a given number of  sites.    Exact enumeration has been carried
out for small lattice animals using a variety of  methods
\cite{edwards:92a,peters:79a,sykes:76a}  but the methods become computationally
prohibitive for large animals. Many techniques have been used to enumerate larger
lattice animals including various Monte Carlo growth schemes \cite{edwards:92a,
lam:96a, stauffer:78a, leath:76a}, a constant fugacity Monte Carlo method
\cite{redner:81a}, an incomplete enumeration method \cite{lam:86a}  and  reaction
limited cluster-cluster aggregation \cite{ball:96a}.

In the following paper we describe an improvement of a method proposed by one of the
authors \cite{care:97a} which was based on an extension of the scheme proposed by
Rosenbluth and Rosenbluth  \cite{rosenbluth:55a} for enumerating self avoiding polymer
chains.  The central problem in using the Rosenbluth scheme for lattice animal
enumeration is calculating the degeneracy of the clusters which are generated. In the
method proposed by Care,  the cluster growth was modified in a way which forced the
degeneracy to be $N!$ where $N$ is the number of sites occupied by the lattice animal.
However the resulting algorithm was fairly complicated to implement. An alternative
method of correcting for the degeneracy had been proposed by Pratt \cite{pratt:82a}.
In this latter scheme the correcting weight is more complicated to determine and must
be recalculated at each stage of the cluster growth if results are sought at each
cluster size. However the Pratt scheme does not require any restriction on the growth
of the cluster.

In this paper we show that there are a class of Rosenbluth like algorithms which yield
a degeneracy of $N$ and which are straightforward to implement.  The method provides
an estimate of the number of lattice animals and can also yield estimates of any other
desired properties of the animals such as their radius of gyration or perimeter
multiplicities \cite{edwards:92a}.  We describe and justify the algorithm in Section
\ref{algorithm} and present results to illustrate the use of the method in Section
\ref{results}.  Conclusions are given in Section \ref{sec:concs}

\section{Algorithm}
\label{algorithm}

Any algorithm, suitable for the purpose of the enumeration of lattice animals using
the Rosenbluth Monte Carlo approach, must satisfy two important criteria.  First of
all it has to be ergodic.  That is to say, the algorithm should have a non zero
probability of sampling any given cluster shape.  The second criteria relates to the
degeneracy that is associated with each cluster and requires this to be determinable.
This degeneracy arises from the number of different ways that the same cluster shape
can be constructed by the algorithm.  While it is easy to devise  methods of growing
clusters that meet the first requirement, the second condition is more difficult to
satisfy.  For many simple algorithms the calculation of the degeneracy, for every
cluster, can be a more complex problem than the original task of enumerating the
number of lattice animals.

In the original Rosenbluth Monte Carlo approach of Care \cite{care:97a}, this
difficulty was overcome by ensuring that the degeneracy for all clusters of size $N$
was the same and equal to $N!$.  However, to achieve this result the algorithm had to
employ a somewhat elaborate procedure.  This made the implementation of the method
rather complicated, as well as limiting its possible extension to enumeration of other
type of clusters.  Here we shall consider an alternative algorithm, which while
satisfying both of the above criteria, is considerably simpler than the algorithm
proposed by Care.  In Section \ref{basic_algorithm} we describe the algorithm in its
most basic form, before proving in Section \ref{ergodicity} that the ergodicity and
the degeneracy requirements are both met.  In Section \ref{refined_algorithm} we
demonstrate how the basic algorithm can be further refined to improve its efficiency.

\subsection{Basic Algorithm}
\label{basic_algorithm}

Having chosen a suitable lattice on which the clusters are to be grown (square and
simple cubic lattices were used in this study for 2D and 3D systems, respectively), a
probability $p$ of acceptance and $q = (1-p)$ of rejecting sites is specified.
Although in principle any value of p between 0 and 1 can be selected, the efficiency
of the sampling process is largely dependent on a careful choice of this value, as
will be discussed later. In addition, an ordered list of all neighbours of a site on
the lattice is made. For example, for a 2D square lattice this might read (right,
down, left, up).  While the order initially chosen is arbitrary, it is essential that
this remains the same throughout a given run.  In the basic algorithm, once chosen,
the probability $p$ remains fixed during the Monte Carlo sampling procedure. However
in Section \ref{refined_algorithm} the effect of  relaxing this requirement is
discussed.

We construct an ensemble of $N_{E}$ clusters and for each of these calculate a weight
factor which we subsequently use to calculate weighted averages of various cluster
properties.  For a property $O$ of the clusters, the weighted average is defined as
\begin{equation}
<O>_W = \frac{1}{N_E}\sum_{\alpha = 1}^{N_E} W_{\alpha} O_{\alpha} \label{eq:wav}
\end{equation}
The weight associated with cluster $\alpha$ with $N$ sites is defined to be
$W_{\alpha} = 1/ (d_{N} P_{\alpha})$ where $P_{\alpha}$ is the normalised probability
of growing the cluster and $d_N$ is a degeneracy equal to the number of ways of
growing a particular cluster shape.  It can be  shown \cite{care:97a} that the
weighted average can be used to estimate the number, $c_N$, of lattice animals of size
$N$ and other properties such as the average radius of gyration $\overline{R_N^{2}}$:-
\begin{eqnarray}
E[<1>_W] &  = & c_N  \label{eq:mn} \\ E[<R^2_{\nu}>_W] & = & \sum_{ \{\nu = 1\}
}^{c_N}R_{N\nu}^2   = c_N \overline{R_N^{2}}
\end{eqnarray}

During the growth of each cluster we maintain a record of the sites which have been
occupied, the sites which have been rejected and a \lq last-in-first-out stack' of
sites which is maintained according to the rules described below. Each cluster is
grown as follows
\begin{enumerate}
\item Starting from an initial position, the neighbours of this site are examined one
at a time according to the list specified above.  An adjacent site is accepted with a
probability p or else is rejected.
\item If  the adjacent site is rejected, a note of this is made and the next neighbour in the
list is considered.

\item If on the other hand it is accepted, then this becomes the current site and its
position is added to top of a stack, as well as to a list of accepted sites.  The
examination of the sites is now resumed for the neighbours of this newly accepted
site.  Once again this is done in the strict order which was agreed at the start of
the algorithm.

\item Sites that have already been accepted or rejected are no longer available for
examination. Thus, if such a site is encountered, it is ignored and the examination is
moved on to the next eligible neighbour in the list.

\item \label{stack_drop} If at any stage the current site has no more neighbours left,
that is all its adjacent sites are already accepted or rejected,  then the current
position is moved back by one to the previous location.  This will  be the position
below the current one in the stack.  The current position is removed from the top of
the stack, though not from the list of accepted sites.

\item \label{stop} The algorithm stops for one of the following two reasons.  If ever
the number of accepted sites reaches N, then the algorithm is immediately terminated.
In this case a cluster of size N is successfully produced.  Note that unlike some of
the other common cluster growth algorithms \cite{leath:76a}, it is not necessary here
for every neighbour of the generated cluster to be rejected.  Some of these might
still be unexamined before the algorithm terminates.  The second way in which the
algorithm stops is when it fails to produce a cluster of size $N$.  In this case, the
number of accepted sites will be $M<N$, with all the neighbours of these $M$ sites
already having been rejected, leaving no eligible sites left for further examination.
From step \ref{stack_drop}, it is clear that in cases such as this, the current
position would have returned to the starting location.

\item The probability of producing a cluster of size N, in a manner involving $r$
rejections, is simply $p^{(N-1)} q^{r}$.  Hence the weight, $W_{\alpha}$, associated
with the growth of the cluster is given by

\begin{equation}
W_{\alpha}= 1/(d_N p^{(1-N)} (1-p)^{r}) \label{eq:weight1}
\end{equation}
where the degeneracy, $d_N$, is shown below to be exactly $N$. Failed attempts have a
zero weight associated with them.  However they must be included in the weighted
average of equation (\ref{eq:wav}).

\item During the growth of a cluster of size $N$, we may also collect data for all the
clusters of size $M$ where $M \leq N$. It must be remembered that the weights for
these smaller clusters must be calculated with a degeneracy of $M$.

\end{enumerate}

\begin{figure}[t]

\resizebox{1.0 \linewidth}{!}{\includegraphics{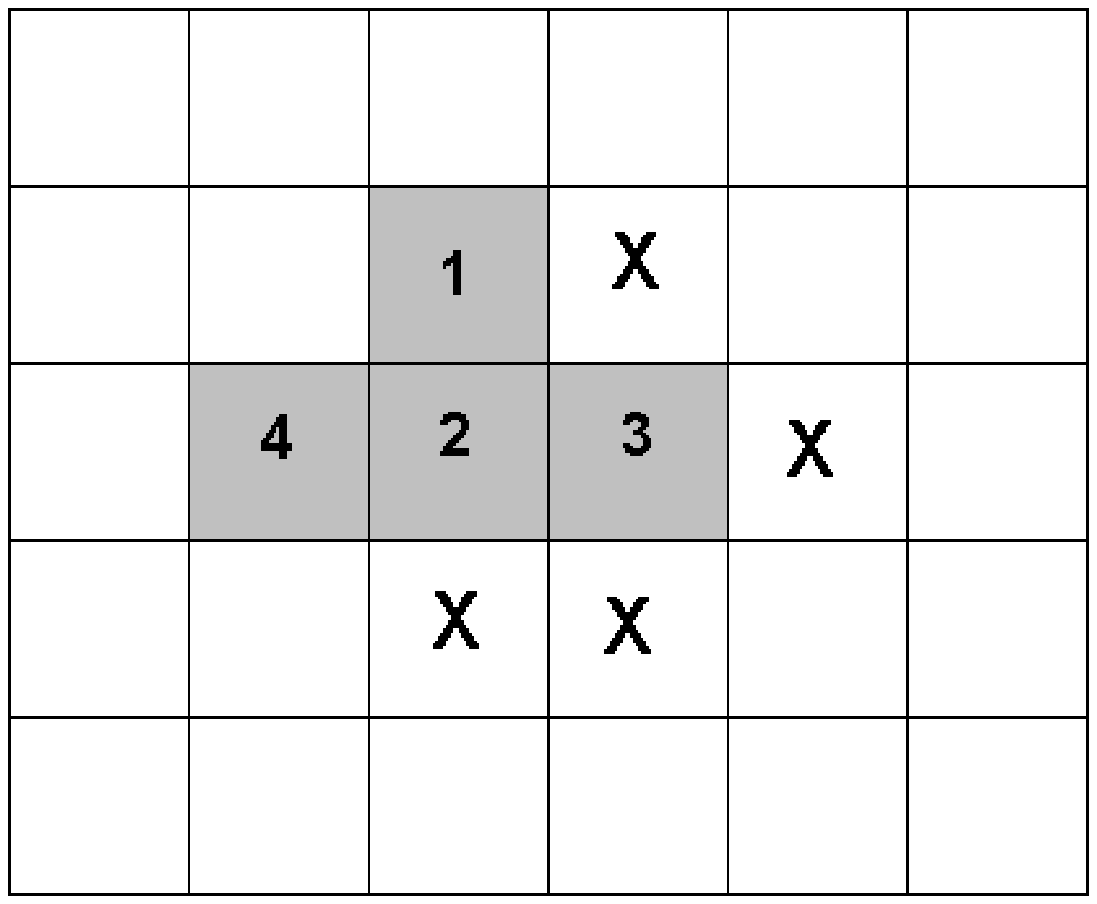}}
\caption{Sequence of
accepted sites leading to a cluster of size $N=4$.  The sites examined but rejected
along the way are indicated by X.  In our notation this sequence can be represented by
\{0,1,1,0,0,0,1\}} \label{fig:example}
\end{figure}

A specific example is helpful in demonstrating the algorithm.  Figure
\ref{fig:example} displays a successful attempt in forming a cluster of size $N=4$, on
a square lattice.  The order in which the neighbours were examined was chosen to be
right, down, left and up.  Let us now consider various steps involved in construction
of this cluster in detail.  Beginning from the initial position labelled cell one, the
adjacent site to the right of this position is examined.  In this case the site is
rejected and the current position remains on the cell one. Such rejected cells are
indicated by the letter X.  The next neighbour in the list is the one below, labelled
cell two.  As it happens this is accepted.  Thus, the current position moves to this
site and its position is added to the top of the stack, ahead of the position of cell
one.  The process of examining the neighbours is resumed for sites adjacent to cell
two.  Once again, following the strict order in the list, the site labelled three to
the right of current position is considered first.  This is also accepted and as
before is placed at the top of the stack.  At this stage the stack contains the
positions of cells three, two and one, in that order.  The current position is now
cell three.

The site to the right of this, followed by the one below, are tested and both rejected
in succession.  Since both the neighbours to the left ( {\em ie} cell one) and the one
above have already been considered, the current position has no more eligible
neighbours left to test.  Therefore, following step \ref{stack_drop} above, site three
is removed from the stack.  This leaves the position of cell two at the top of the
stack, making this the current position again.  The cell two has two neighbours, the
adjacent sites below and to the left, which are still unexamined.  Of these, according
to our agreed list, the site below takes precedent, but as shown in Figure
\ref{fig:example} this is rejected.  Current position remains on the cell two and the
neighbouring site (cell labelled four) to the left of this position is tested.  As it
happens this is accepted.  A cluster of the desired size $N=4$ is achieved, bringing
this particular attempt to a successful end.

For the subsequent discussion, it is useful to represent a sequence of acceptance and
rejections by a series of 1 and 0.  Thus, for the case shown in Figure
\ref{fig:example} we have \{0,1,1,0,0,0,1\}.  Note that at any stage throughout a
series, the position of the current site and that of the neighbour to be examined,
relative to the starting cell, are entirely specified by the decisions that have been
made so far.  In other words, given a sequence of one and zeros we can determine
precisely the shape of the cluster that was constructed. This is only possible because
of the manner in which the neighbours of the current position are always tested in a
strict pre-defined order.  For an algorithm that considers the neighbouring sites at
random, the same will clearly not be true.

The procedure described above needs to be repeated a large number of times, to obtain
the weights for the ensemble average defined in equation (\ref{eq:wav}).  In
particular, using equation \ref{eq:mn}, the number of lattice animals of size $N$ can
now be determined.

\subsection{Ergodicity and degeneracy of the algorithm }
\label{ergodicity}

Let us now discuss the issue of the ergodicity of the algorithm.  We wish to see
whether, starting from any particular site on a given cluster, a series of acceptance
and rejections (1 and 0) can always be determined which leads to that cluster shape.
We stress that we are not concerned about how probable such a sequence is likely to
be, but merely that it exists.  We can attempt to construct such a sequence by
following the same rules as our algorithm described above, with one exception; we
accept and reject each examined site according to whether it forms part of the target
cluster shape or not.   Obviously, in the original algorithm, each such move has a non
zero chance of occurring, provided $p$ is not set to zero or one.  Since we only
accept sites that belong to the cluster in question, it follows that if the sequence
is successful then we would achieve the desired cluster shape.  However, we might
argue that for some choice of target cluster and starting position,  a series started
in this manner will always terminate prematurely.  That is to say, it will inevitably
lead to a failure, with only part of the required cluster having been constructed.
Now, it is easy to see that this cannot be true.  If the series fails, it implies that
all the neighbouring sites of the sub-cluster formed so far are rejected.  However,
the rest of the cluster must be connected to this sub-cluster at some point.  Hence,
at very least, one neighbouring site of the sub-cluster must be part of the full
cluster and could not have been rejected.  Starting from any of the sites belonging to
a cluster then, it is always possible to write down a sequence of one and zeros that
will result in the formation of that cluster.  Similarly, considering every starting
point on a cluster of size $N$, another implication of the above result is that the
corresponding cluster shape can be generated in a minimum of  at least $N$ distinct
ways.

Next, we shall show that the degeneracy of a cluster of size $N$ in our algorithm is
in fact exactly $N$ (unlike the original algorithm of Care \cite{care:97a} which has a
degeneracy of $N!$). Let us suppose that starting from a particular site on a given
target cluster shape, our algorithm has two distinct ways of forming this cluster.
Associated with each of these, a series of one and zeros can be written down, in the
same manner as that indicated above.  The two ways of constructing the cluster must
necessarily begin to differ from each other at some stage along the sequence, where we
will have a 1 in one case and a 0 in the other.  Now since up to this point the two
series are identical, the site being examined at this stage will be the same for both
cases.  This is rejected in one sequence (hence 0) whereas it is accepted in the other
(hence 1).  It immediately follows that these two differing ways of constructing the
cluster cannot result in the same shape.  Using this result, together with previous
one regarding the ergodicity of the algorithm, we are lead to conclude that, starting
from a given site on a cluster, the algorithm has one and only one way of constructing
the cluster.  Hence, for a cluster of  size $N$, the degeneracy is simply $N$.

\subsection{Refined algorithms}
\label{refined_algorithm}
\subsubsection{Adjacent site stack}
\label{adjacentsitestack} During the growth of the cluster a stack can be constructed
of all the sites which are adjacent to the cluster and still available for growth.
When a new site is added to the cluster, its neighbours are inspected in the
predetermined sequence and any available ones are added to the top of this stack.
(Note that this stack differs from that discussed in Section (\ref{basic_algorithm})).
The choice of site to be occupied can be made from all the adjacent sites in a single
Monte Carlo decision. Thus, if we consider the underlying process in the method
described above, at each step there is a probability $p$ of the site being accepted
and a probability $q=1-p$ of the site being rejected. We therefore need to generate a
random number with the same distribution as the number of attempts needed to obtain an
acceptance. The probability of making $k$ attempts of which only the last is
successful, is
\begin{equation}
p_k = q^{k-1} p
\end{equation}
where $1\leq k < \infty$ and $\sum_{k=1}^{\infty} p_k = 1$.  In order to sample from
this distribution we note that the associated cumulative distribution,  $C_m$,  is
given by
\begin{equation}
C_m = \sum_{k=1}^{m} q^{k-1} (1 - q) = 1 - q^m
\end{equation}
Hence if we generate a random number, $\eta$,  uniformly distributed in the range $0 <
\eta < 1$,  then a number $m$ given by
\begin{equation}
m  = {\rm Int}\left[ \frac{\ln(\eta)}{\ln(q)} + 1 \right] \label{equ:m}
\end{equation}
will have been drawn from the required distribution.  Thus we generate the number $m$
according to equation (\ref{equ:m}) and use this to determine which site on the stack
is selected, with $m=1$ corresponding to the site at the top of the stack.  If $m >
N_{adj}$, where $N_{adj}$ is the number of available adjacent sites, the cluster
growth is terminated as explained in step (\ref{stop}) in Section
\ref{basic_algorithm}.  All the adjacent sites lying above the chosen site in the
stack are transferred into the list of rejected sites. The list of adjacent sites is
then adjusted to include the new available sites adjacent to the recently accepted
site. As before, it is crucial that these are added to the top of the list in the
strict predefined order.
\subsubsection{Variable probability}
\label{variableprob} An apparent disadvantage of the methods so far described is that
with fixed choice of probability, $p$, occasions  arise when a cluster growth will
terminate before reaching a cluster of size $N$,  simply because the Monte Carlo choice
rejected all the neighbouring sites. This problem can be overcome if the value of $p$ is
allowed to vary as the cluster grows. The simplest method is to determine the number,
$N_{adj}$, of available adjacent sites at each point in the cluster growth and select one
of these sites with uniform probability. This effectively makes $p = 1 /N_{adj}$ and
thereby increases the chances of growing a cluster of size $N$. Note that it is still
possible for a cluster growth to become blocked. This happens when the chosen site is the
one at the bottom of the current eligible neighbours list, thus causing all the other
neighbouring sites in the list to be rejected in one step. If the newly accepted site has
itself no unexamined neighbours to add to the list, the algorithm terminated prematurely.
Modified in the manner described above the weight associated with a cluster is now
\begin{equation}
W_{\alpha} = \frac{ \Pi_{i=1}^{N} N_{adj}^{i}}{N}
\end{equation}
rather than the expression given in equation (\ref{eq:weight1}).

However, when this variable probability method was tested it was found that although
it reduced the number of rejected clusters, it was inefficient at sampling the space
of possible clusters when compared with method described in section
(\ref{adjacentsitestack}).  This inefficiency was measured by comparison of the
standard deviation in the estimated cluster number for any given number of clusters in
the sampling ensemble. It is thought that the inefficiency of the variable probability
method arises because it gives too much weight to sites lower in the stack, yielding
many non-representative clusters. It is possible that this problem could be overcome
by using a non-uniform sampling distribution ({\em cf} \cite{care:97a}) but this was
not tested in this work and the method described in (\ref{adjacentsitestack}) was used
to obtain the results described in Section (\ref{results}) .

\section{Results}
\label{results}

\begin{figure}[t]
\resizebox{0.95 \linewidth}{!}{\includegraphics{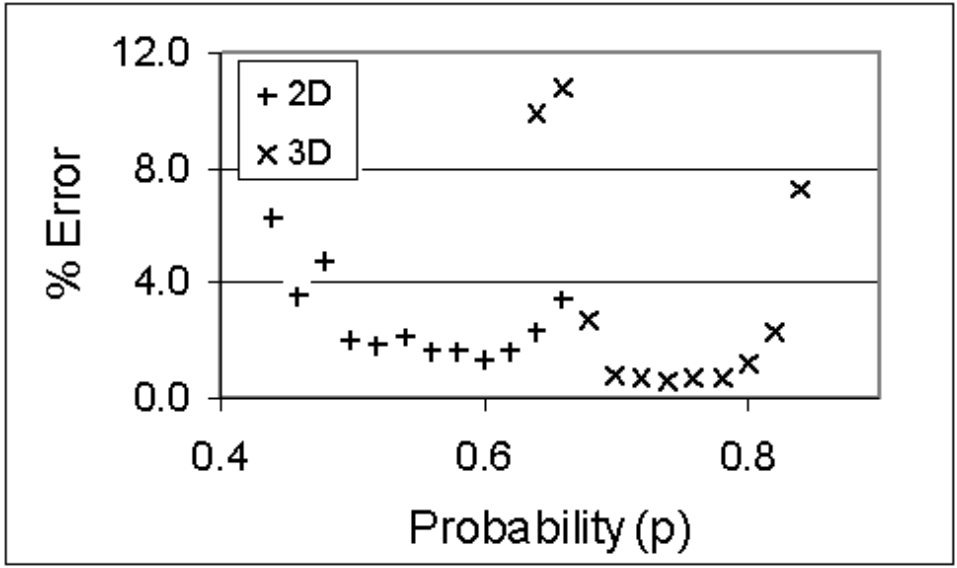}} \caption{Percentage errors
for clusters of size 50} \label{fig:err50}
\end{figure}

In order to test the algorithm described in Section (\ref{algorithm}) it was used to
estimate the number of lattice animals on a square 2D lattice and a simple cubic 3D
lattice for which exact results are known up to certain sizes \cite{sykes:76a}. Before
collecting data it was necessary to determine the optimum value of the probability $p$
with which an adjacent site is accepted during the cluster growth.  The effect of
changing $p$ on the estimated error in the number of clusters of size 50 on the 2D and
3D lattices can be seen in Figure \ref{fig:err50}.   It can be seen that there is a
fairly broad range of values of $p$ for which the error is a minimum and a value of $p
= 0.6$ was used to obtain the results described below for the 2D lattice and $0.72$
for the 3D lattice. The distribution of weights is log normal \cite{care:97a} and
becomes highly skewed for large cluster sizes; this is a standard problem with
Rosenbluth methods \cite{batoulis:88a}. The minimum in the error achieved by the
choice of the value of the probability $p$ has the effect of minimising the variance
of the distribution of the weights, $W_{\alpha}$.

In Table \ref{tab:3db}  we present results obtained using the algorithm defined in
section \ref{algorithm} using the adjacent site stack method of section
\ref{refined_algorithm} to enumerate clusters on a simple cubic 3D lattice for
clusters up to size 50. The results were obtained from an ensemble of $2.5 \times 10^7
$ clusters. The data took  3.3 hours to collect on a R5000 Silicon Graphics
workstation using code written in the language C but with no attempt to optimise the
code.  Only $30\%$ of the clusters achieved a size of 50. The results are quoted
together with a standard error, $e^{est}$, calculated by breaking the data into 50
blocks and determining the variance of the block means for each cluster size. If the
number of samples in each block is sufficient, it follows from the central limit
theorem that the sampling distribution of the means should become reasonably
symmetrical. We therefore also quote a {\em skewness}, $\xi$, defined by
\cite{bulmer:65a}
\begin{equation}
\xi = m_3 / m_2^{3 / 2}
\end{equation}
where $m_i$ is the $i^{th}$ moment about the mean of  the sampling distribution. It is
expected that $\xi  \stackrel{<}{\sim} 0.5 $ for a symmetrical distribution and $\xi
> 1 $ for a highly skew distribution.   The statistic $\xi$ should be treated with
some caution since it is likely to be subject to considerable error because it
involves the calculation of a third moment from a limited number of data points.

\begin{figure}[t]
\resizebox{0.95 \linewidth}{!}{\includegraphics{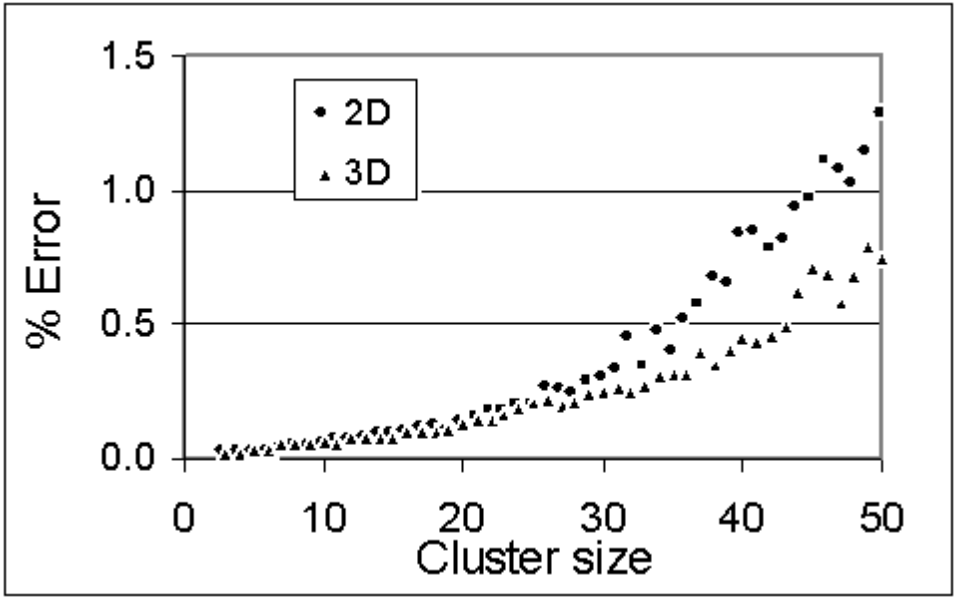}} \label{fig:errN}
\caption{Variation of percentage error with cluster size.}
\end{figure}

Exact results are known for clusters up to size 13 \cite{lam:96a} and in the table we
quote the values for the quantity $\chi$  defined by
\begin{equation}
\chi_M =|\frac{c_M^{exact} - c_M^{est} } {c_M^{exact} e_M^{est}  } |
\end{equation}
where $c_M$ is the number of clusters of size $M$ and it can be seen that all the
values of $\chi$ are $O(1)$. Hence we assume that $e^{est}$ is an acceptable method of
estimating the error in the method. However it is likely that the $e^{est}$ will
underestimate the true error if the distribution becomes more skew. We also quote in
Table \ref{tab:3db} the values of $c_N$ calculated by Lam \cite{lam:96a} using a Monte
Carlo incomplete enumeration method together with the  error estimates reported for
this method.

In Table \ref{tab:2db} we quote data collected from a square two dimensional lattice
by collecting data from $2.5 \times 10^7$ clusters up to size 50. This data only took
1.45 hours to collect but only $2\%$ of the clusters achieved a size of 50. Comparison
is given with exact results \cite{sykes:76a} up to clusters of size 19. The rate of
growth of errors for the two and three dimensional data is shown in Figure
\ref{fig:errN} and it can be seen that the errors associated with the method diverge
are beginning to diverge quite rapidly above clusters of size 50. This behaviour is to
be expected with a technique which is based on sampling from a log normal
distribution. In the previous paper \cite{care:97a} equivalent results were obtained
for clusters up to size 30 with approximately the same sample size. The improvement up
to clusters of size 50 obtained by the new method arises because the weight associated
with clusters of  a certain size is generated from roughly half as many random
numbers. This effectively halves the standard deviation of the log normal distribution
of the weights and allows larger clusters to be sampled before the method becomes
unusable.

\section{Conclusions}
\label{sec:concs} We have described a simple Rosenbluth algorithm for the Monte Carlo
enumeration of lattice animals and clusters which can be applied to any lattice
topology. A merit of the scheme is that for thermal systems it may be easily adapted
to include Boltzmann weightings following, for example, the arguments used by Siepmann
{\em at al} \cite{siepmann:92a} in the development of the configurational bias
technique. Similarly, the method can be applied to calculation of the averaged
properties of a cluster of a given size, in the site percolation problem.  In this
case we have
\begin{equation}
<O>=\frac{<(1-P)^{t}O>_{W}}{<(1-P)^{t}>_{W}}=\frac{\sum_{\alpha=1}^{N_{E}}
W_{\alpha}(1-P)^{t_{\alpha}} O_{\alpha}
}{\sum_{\alpha=1}^{N_{E}}W_{\alpha}(1-P)^{t_{\alpha}}}
\end{equation}
where $P$ is the probability of site occupation in the percolation problem of interest
and $t_{\alpha}$ the number of perimeter sites \cite{stauffer:92a} of the cluster
$\alpha$. Preliminary results also indicate that the method may be useful in the study
of the adsorption of clusters onto solid surfaces. A possible numerical limitation of
the method arises from the highly skew probability distribution  of Rosenbluth weights
which occurs for large cluster sizes. However the method presented in this work  is
able to work to considerably higher cluster sizes than the one described in
\cite{care:97a} before this becomes a problem.

\renewcommand{\baselinestretch}{1}

\bibliography{cmcpre99}

\begin{thebibliography}{10}
\providecommand*{\bibinfo}[2]{#2}
\providecommand*{\eprint}[1]{#1}
\providecommand*{\url}[1]{#1}
\bibitem{jacucci:83a}
\bibinfo{author}{G.~Jacucci}, \bibinfo{author}{A.~Perini}, and
  \bibinfo{author}{G.~Martin}, \bibinfo{journal}{J Phys A:Math and Gen}
  \bibinfo{volume}{\textbf{16}}, \bibinfo{pages}{369} (\bibinfo{date}{1983}).
\bibitem{edwards:92a}
\bibinfo{author}{B.~F. Edwards}, \bibinfo{author}{M.~F. Gyure}, and
  \bibinfo{author}{M.~Ferer}, \bibinfo{journal}{Phys Rev A}
  \bibinfo{volume}{\textbf{46}}, \bibinfo{pages}{6252} (\bibinfo{date}{1992}).
\bibitem{ball:96a}
\bibinfo{author}{R.~C. Ball} and \bibinfo{author}{J.~R. Lee},
  \bibinfo{journal}{J Phys I France} \bibinfo{volume}{\textbf{6}},
  \bibinfo{pages}{357} (\bibinfo{date}{1996}).
\bibitem{peters:79a}
\bibinfo{author}{H.~P. Peters}, \bibinfo{author}{D.~Stauffer},
  \bibinfo{author}{H.~P. H{\"{o}}lters}, and \bibinfo{author}{K.~Loewenich},
  \bibinfo{journal}{Z Physik B} \bibinfo{volume}{\textbf{34}},
  \bibinfo{pages}{339} (\bibinfo{date}{1979}).
\bibitem{sykes:76a}
\bibinfo{author}{M.~F. Sykes} and \bibinfo{author}{M.~Glen},
  \bibinfo{journal}{J Phys A: Math Gen} \bibinfo{volume}{\textbf{9}},
  \bibinfo{pages}{87} (\bibinfo{date}{1976}).
\bibitem{lam:96a}
\bibinfo{author}{P.~M. Lam} and \bibinfo{author}{F.~Family},
  \bibinfo{journal}{Physica A} \bibinfo{volume}{\textbf{231}},
  \bibinfo{pages}{369} (\bibinfo{date}{1996}).
\bibitem{stauffer:78a}
\bibinfo{author}{D.~Stauffer}, \bibinfo{journal}{Phys Rev Lett}
  \bibinfo{volume}{\textbf{41}}, \bibinfo{pages}{1333} (\bibinfo{date}{1978}).
\bibitem{leath:76a}
\bibinfo{author}{P.~L. Leath}, \bibinfo{journal}{Phys Rev Lett}
  \bibinfo{volume}{\textbf{36}}, \bibinfo{pages}{921} (\bibinfo{date}{1976}).
\bibitem{redner:81a}
\bibinfo{author}{S.~Redner} and \bibinfo{author}{P.~J. Reynolds},
  \bibinfo{journal}{J Phys A: Math and Gen} \bibinfo{volume}{\textbf{14}},
  \bibinfo{pages}{2679} (\bibinfo{date}{1981}).
\bibitem{lam:86a}
\bibinfo{author}{P.~M. Lam}, \bibinfo{journal}{Phys Rev A}
  \bibinfo{volume}{\textbf{34}}, \bibinfo{pages}{2339} (\bibinfo{date}{1986}).
\bibitem{care:97a}
\bibinfo{author}{C.~M. Care}, \bibinfo{journal}{Phys Rev E}
  \bibinfo{volume}{\textbf{57}}, \bibinfo{pages}{1181} (\bibinfo{date}{1997}).
\bibitem{rosenbluth:55a}
\bibinfo{author}{M.~N. Rosenbluth} and \bibinfo{author}{A.~W. Rosenbluth},
  \bibinfo{journal}{J Chem Phys} \bibinfo{volume}{\textbf{23}},
  \bibinfo{pages}{356} (\bibinfo{date}{1955}).
\bibitem{pratt:82a}
\bibinfo{author}{L.~Pratt}, \bibinfo{journal}{J Chem Phys}
  \bibinfo{volume}{\textbf{77}}, \bibinfo{pages}{979} (\bibinfo{date}{1982}).
\bibitem{batoulis:88a}
\bibinfo{author}{J.~Batoulis} and \bibinfo{author}{K.~Kremer},
  \bibinfo{journal}{J Phys A: Math Gen} \bibinfo{volume}{\textbf{21}},
  \bibinfo{pages}{127} (\bibinfo{date}{1988}).
\bibitem{bulmer:65a}
\bibinfo{author}{M.~G. Bulmer}, \bibinfo{title}{\emph{Principles of
  Statistics}} (\bibinfo{publisher}{Oliver and Boyd}, London,
  \bibinfo{year}{1965}).
\bibitem{siepmann:92a}
\bibinfo{author}{J.~I. Siepmann} and \bibinfo{author}{D.~Frenkel},
  \bibinfo{journal}{Mol Phys} \bibinfo{volume}{\textbf{75}},
  \bibinfo{pages}{59} (\bibinfo{date}{1992}).
\bibitem{stauffer:92a}
\bibinfo{author}{D.~Stauffer}, \bibinfo{author}{A.~Aharony}, and
  \bibinfo{author}{Taylor}, \bibinfo{title}{\emph{Introduction to percolation
  theory}} (\bibinfo{publisher}{Taylor and Francis}, \bibinfo{year}{1992}).

\end{thebibliography}
\bibliographystyle{revtex}

\listoffigures

\clearpage

\begin{table}
\centering
\begin{tabular}{|c||c|r|c|c|c|c|c|r|}
\hline $N$      & Rosenbluth & \multicolumn{1}{c|}{Exact} & Lam \cite{lam:96a}&
$e^{est}$         & True      & Lam \cite{lam:96a}  & $\chi$ &
\multicolumn{1}{c|}{$\xi$} \\ & estimate         &  \multicolumn{1}{c|}{value}
&estimate        & \% error & \% error & \% error &        &    \\

\hline \hline

2 &3.000$\times 10^{0}$ & 3  &   &    &   &        & &\\

3 & 1.499$\times 10^{1}$ & 15 &  &   &  & &  & \\

4 & 8.600$\times 10^{1}$ & 86 & 8.594$\times 10^{1}$ & 0.03 & 0.00 & 0.51 & 0.18 &
0.07\\

5 & 5.339$\times 10^{2}$ &    534 & 5.321$\times 10^{2}$ & 0.03 & 0.02 & 0.54 & 0.77 &
0.00\\

6 & 3.483$\times 10^{3}$ & 3 481       & 3.475$\times 10^{3}$ & 0.04 & 0.05 & 0.58 &
1.30& 0.14\\

7 & 2.351$\times 10^{4}$ & 23 502 & 2.353$\times 10^{4}$ & 0.05 & 0.02 & 0.63 & 0.42 &
0.14\\

8 & 1.630$\times 10^{5}$ &162 913      & 1.631$\times 10^{5}$ & 0.05  & 0.03& 0.65 &
0.58 & 0.73\\

9 & 1.153$\times 10^{6}$ &1 152 870    & 1.155$\times 10^{6}$ & 0.06 & 0.03 & 0.73 &
0.50& 0.62\\

10 &8.302$\times 10^{6}$&8 294 738    & 8.291$\times 10^{6}$ & 0.06 & 0.09 & 0.86
&1.40 & 0.16\\

11 &6.054$\times 10^{7}$&60 494 540  & 6.042$\times 10^{7}$ & 0.06 & 0.08 & 0.87 &
1.29 & 0.50\\

12 &4.464$\times 10^{8}$&446 205 905 & 4.442$\times 10^{8}$ &0.07 & 0.05 & 0.87  &
0.70 & 0.12 \\

13 &3.326$\times 10^{9}$&3 322 769 129 &3.291$\times 10^{9}$&0.08&0.11 & 0.97& 1.34  &
0.48\\

14 & 2.496$\times 10^{10}$& &2.461$\times 10^{10}$&0.07 & &1.09 & &0.35\\

15 & 1.887$\times 10^{11}$& &1.862$\times 10^{11}$&0.07 &   &1.16   & &-0.10
\\

16 & 1.436$\times 10^{12}$& &1.416$\times 10^{12}$&0.10 &   &1.22   & &0.25\\

17 & 1.098$\times 10^{13}$& &1.082$\times 10^{13}$&0.10 &   &1.27   & &-0.03\\

18 & 8.448$\times 10^{13}$& &8.329$\times 10^{13}$&0.09 &   &1.37   & &0.12\\

19 & 6.520$\times 10^{14}$& &6.446$\times 10^{14}$&0.11 &   &1.38   & &0.20\\

20 & 5.048$\times 10^{15}$& &5.002$\times 10^{15}$&0.13 &   &1.41   & &-0.07\\

21 & 3.929$\times 10^{16}$& &3.897$\times 10^{16}$&0.14 &   &1.47   & &-0.21\\

22 &3.063$\times 10^{17}$& &3.052$\times 10^{17}$&0.14 &   &1.49   & &-0.42\\

23 & 2.399$\times 10^{18}$& &2.391$\times 10^{18}$&0.16 &   &1.61   & &-0.11\\

24 & 1.882$\times 10^{19}$& &1.877$\times 10^{19}$&0.19 &   &1.68   & &0.16\\

25 & 1.485$\times 10^{20}$& &1.480$\times 10^{20}$&0.21 &   &1.70   & &-0.02\\

26 & 1.169$\times 10^{21}$& &1.168$\times 10^{21}$&0.21 &   &1.75   & &-0.11\\

27 & 9.214$\times 10^{21}$& &9.209$\times 10^{21}$&0.20 &   &1.81   & &0.06\\

28 & 7.316$\times 10^{22}$& &7.290$\times 10^{22}$&0.21 &   &1.88   & &0.18\\

29 & 5.790$\times 10^{23}$& &5.786$\times 10^{23}$&0.24 &   &1.96   & &-0.12\\

30 &4.600$\times 10^{24}$& &4.610$\times 10^{24}$&0.25 &   &2.01   & &0.44\\

\hline
\end{tabular}
\caption{Table continued on next page} \label{tab:3da}
\end{table}

\addtocounter{table}{-1}

\clearpage

\begin{table}
\centering
\begin{tabular}{|c||c|r|c|c|c|c|c|r|}
\hline $N$      & Rosenbluth & \multicolumn{1}{c|}{Exact} & Lam \cite{lam:96a}&
$e^{est}$         & True      & Lam \cite{lam:96a}  & $\chi$ &
\multicolumn{1}{c|}{$\xi$} \\ & estimate         &  \multicolumn{1}{c|}{value}
&estimate        & \% error & \% error & \% error &        &    \\

\hline \hline

31 &3.674$\times 10^{25}$& &&0.26 &   &  & &-0.28\\

32 &2.929$\times 10^{26}$& &&0.25 &   &   & &0.26\\

33 &2.342$\times 10^{27}$& &&0.27 &   & & &0.54\\

34 &1.872$\times 10^{28}$& &&0.31 &   &  & &0.46\\

35 &1.501$\times 10^{29}$& &&0.31 &   &   & &-0.32\\

36 &1.199$\times 10^{30}$& &&0.32 &   &   & &0.33\\

37 &9.631$\times 10^{30}$& &&0.39&   & & &1.08\\

38 &7.691$\times 10^{31}$& &&0.35 &   &   & &0.18\\

39 &6.203$\times 10^{32}$& &&0.40 &   &  & &0.27\\

40 &4.984$\times 10^{33}$& &&0.45 &   &   & &0.54\\

41 &3.999$\times 10^{34}$& &&0.43 &   &   & &0.35\\

42 &3.205$\times 10^{35}$& &&0.46 &   &  & &0.23\\

43 &2.605$\times 10^{36}$& &&0.49 &   &   & &0.35\\

44 &2.100$\times 10^{37}$& &&0.62 &   &   & &2.32\\

45 &1.684$\times 10^{38}$& &&0.71 &   &  & &0.43\\

46 &1.353$\times 10^{39}$& &&0.69 &   & & &0.65\\

47 &1.087$\times 10^{40}$& &&0.58 &   &  & &0.36\\

48 &8.892$\times 10^{40}$& &&0.68 &   &  & &0.53\\

49 &7.223$\times 10^{41}$& &&0.79 &   &  & &0.02\\

50 &5.789$\times 10^{42}$& &&0.75 &   &   & &0.78\\

 \hline
\end{tabular}
\caption{Continued:-   Degenerate Rosenbluth estimate of the number of  lattice
animals of size $N$ on a three dimensional square lattice using $2.5 \times 10^7$
sample clusters, each grown to $N = 50$ with $p = 0.72$; exact values from
\protect\cite{lam:96a}; estimated values and associated errors from incomplete
enumeration method of Lam \protect\cite{lam:96a}; calculation of error estimate
described in text; \protect\lq true' error is fractional difference of Rosenbluth
estimate and exact value; $\chi$ and $\xi$ are defined in the text.} \label{tab:3db}
\end{table}

\clearpage

\begin{table}
\centering
\begin{tabular}{|c||c|r|c|c|c|r|}
\hline $N$ & Rosenbluth &  \multicolumn{1}{c|}{Exact} &    $e^{est}$   & True      &
$\chi$ &  \multicolumn{1}{c|}{$\xi$} \\
     &   estimate   &   \multicolumn{1}{c|}{value}  &   \% error    &  \%error  &         &     \\
\hline \hline

2  & 1.999$\times  10^{0 }$ &  2    &     &  &   &  \\

3  & 6.000$\times  10^{0 }$ &  6    & 0.02  & 0.01  & 0.22  & -0.48\\

4  & 1.900$\times 10^{1 }$ &  19    & 0.03  & 0.00      & 0.00  & -0.65\\

5  & 6.300$\times   10^{1 }$ & 63   & 0.03  & 0.01      & 0.31   & 0.14\\

6  & 2.160$\times   10^{2 }$ &216   & 0.03  & 0.00      & 0.00  & 0.36\\

7  & 7.601$\times 10^{2 }$ & 760     & 0.04 & 0.02   & 0.43     & -0.27\\

8  & 2.724$\times   10^{3 }$ & 2 725 & 0.04  & 0.03  & 0.60  & 0.08 \\

9  & 9.903$\times   10^{3 }$ & 9 910 & 0.05  & 0.07 & 1.48  & -0.14 \\

10 & 3.644$\times 10^{4}$ &36 446    & 0.05 & 0.01   & 0.21 & 0.10\\

11 & 1.352$\times    10^{5}$ &135 268 & 0.06   & 0.04 & 0.69  & 0.09\\

12 & 5.056$\times    10^{5}$ &505 861  & 0.07   & 0.04 & 0.66 & -0.04\\

13 & 1.903$\times    10^{6}$ &1 903 890 & 0.08   & 0.04  & 0.51  & -0.24\\

14 & 7.205$\times    10^{6}$ &7 204 874  & 0.09   & 0.01 & 0.06  & -0.13\\

15 & 2.741$\times   10^{7 }$ & 27 394 666   &0.09   & 0.05      & 0.49      & -0.33 \\

16 & 1.046$\times   10^{8 }$ &104 592 937   &0.09   & 0.01  & 0.07      & -0.09  \\

17 & 4.009$\times   10^{8 }$& 400 795 844   &0.11   & 0.03  & 0.29      & 0.74  \\

18 & 1.543$\times   10^{9 }$ & 1 540 820 542    &0.12   & 0.13  & 1.09      & 0.44 \\

19 & 5.942$\times   10^{9 }$ & 5 940 738 676     & 0.10 & 0.01  & 0.15      & 0.26 \\

20 & 2.298$\times   10^{10 }$&  & 0.13  &   &   & -0.42 \\

21 & 8.895$\times 10^{10 }$&     & 0.15   &  &   & -0.02 \\

22 & 3.451$\times   10^{11 }$& & 0.17  &   & & 0.62 \\

23 & 1.341$\times   10^{12 }$&  & 0.18  &   &   & 0.61 \\

24 & 5.228$\times 10^{12 }$&    & 0.20  &   &   & 1.61 \\

25 & 2.039$\times 10^{13 }$&    & 0.19  &   &   & -0.04  \\

26 & 7.970$\times 10^{13 }$&   & 0.26 &   &   & -0.05 \\

27 & 3.122$\times 10^{14 }$& & 0.25 &   &   & 0.00 \\

28 & 1.225$\times   10^{15 }$&  & 0.24  &   & & 0.33\\

29 & 4.831$\times 10^{15 }$&  & 0.28  &   &   & 0.20 \\

30 & 1.883$\times 10^{16 }$& & 0.30 & &   & -0.13 \\

\hline \end{tabular} \caption{Table continued on next page} \label{tab:2da}
\end{table}
\addtocounter{table}{-1}

\clearpage

\begin{table}
\centering
\begin{tabular}{|c||c|r|c|c|c|r|}
\hline $N$ & Rosenbluth &  \multicolumn{1}{c|}{Exact} &    $e^{est}$   & True      &
$\chi$ &  \multicolumn{1}{c|}{$\xi$} \\
     &   estimate   &   \multicolumn{1}{c|}{value}  &   \% error    &  \%error  &
     &     \\
\hline \hline

31 & 7.426$\times 10^{16 }$& & 0.33 & &   & 0.97 \\

32 & 2.945$\times 10^{17 }$& & 0.45 & &   & 0.59 \\

33 & 1.160$\times 10^{18 }$& & 0.34 & &   & 0.19 \\

34 & 4.561$\times 10^{18 }$& & 0.47 & &   & 0.44 \\

35 & 1.800$\times 10^{19 }$& & 0.40 & &   & 0.23 \\

36 & 7.121$\times 10^{19 }$& & 0.52 & &   & 0.29 \\

37 & 2.823$\times 10^{20 }$& & 0.57 & &   & 0.67 \\

38 & 1.122$\times 10^{21}$& & 0.67 & &   & -0.03 \\

39 & 4.417$\times 10^{21 }$& & 0.65 & &   & 0.71 \\

40 & 1.763$\times 10^{22 }$& & 0.83 & &   & 1.30 \\

41 & 6.979$\times 10^{22 }$& & 0.84 & &   & 1.02 \\

42 & 2.738$\times 10^{23 }$& & 0.78 & &   & 0.37 \\

43 & 1.088$\times 10^{24 }$& & 0.82 & &   & -0.16 \\

44 & 4.341$\times 10^{24 }$& & 0.93 & &   & 2.12 \\

45 & 1.704$\times 10^{25 }$& & 0.97 & &   & 0.52 \\

46 & 6.802$\times 10^{25 }$& & 1.10 & &   & 0.73 \\

47 & 2.673$\times 10^{26 }$& & 1.07 & &   & 0.41 \\

48 & 1.058$\times 10^{27 }$& & 1.02 & &   & 0.60 \\

49 & 4.209$\times 10^{27 }$& & 1.14 & &   & 0.26 \\

50 & 1.664$\times 10^{28 }$& & 1.28 & &   & 0.29 \\

\hline \end{tabular} \caption{Continued:- Degenerate Rosenbluth estimate of the number
of lattice animals of size $N$ on a two dimensional square lattice using $2.5 \times
10^7$ sample clusters, each grown to $N = 50$ with $p = 0.60$; exact results from
\protect\cite{sykes:76a}; calculation of error estimate described in text; \protect\lq
true' error is fractional difference of Rosenbluth estimate and true value; $\chi$ and
$\xi$ are defined in the text.} \label{tab:2db}
\end{table}

\end{document}